\documentclass[preprint,amsmath,superscriptaddress,showpacs,showkeys,amssymb,aps,apl]{revtex4-1}
\usepackage{graphicx,epsfig,multirow,dcolumn,bm}

\begin{document}
\title{Phase separation and effect of strain on magnetic properties of Mn$_3$Ga$_{1-x}$Sn$_x$C}

\author{E. T. Dias}
\affiliation{Department of Physics, Goa University, Taleigao Plateau, Goa 403206 India}
\author{A. Das}
\affiliation{Solid State Physics, Division, Bhabha Atomic Research Centre, Trombay, Mumbai 400085}
\author{A. Hoser}
\affiliation{Helmholtz-Zentrum Berlin, 14109, Berlin, Germany}
\author{S. Emura}
\affiliation{Institute of Scientific and Industrial Research, Osaka University, Osaka, Japan}
\author{A. K. Nigam}
\affiliation{Tata Institute of Fundamental Research, Dr. Homi Bhabha Road, Colaba, Mumbai 400005, India}
\author{K. R. Priolkar}
\email{krp@unigoa.ac.in}
\affiliation{Department of Physics, Goa University, Taleigao Plateau, Goa 403206 India}

\begin{abstract}
While the unit cell volume of compounds belonging to the Mn$_3$Ga$_{1-x}$Sn$_x$C, (0 $ \le x \le $ 1) series shows a conformity with Vegard's law, their magnetic and magnetocaloric properties behave differently from those of parent compounds Mn$_3$GaC and Mn$_3$SnC. A correlation between the observed magnetic properties and underlying magnetic and local structure suggests that replacing Ga atoms by larger atoms of Sn results in the formation of Ga-rich and Sn-rich clusters. As a result, even though the long range structure appears to be cubic, Mn atoms find themselves in two different local environments. The packing of these two different local structures into a single global structure induces tensile/compressive strains on the Mn$_{6}$C functional  unit and is responsible for the observed magnetic properties across the entire solid solution range.
\end{abstract}
\date{\today}

\maketitle

\section {Introduction}
Doping in a compound is generally used as an effective strategy to improve or even completely change properties of a material.  Doping creates a local positive or negative pressure, known as chemical pressure which alters the interatomic distances, bond angles or even the crystal structure of a material affecting its properties. Even in Mn based antiperovskites of type ABMn$_{3}$ which exhibit a variety of phase transitions and technologically significant phenomena, doping has been used quite effectively to improve the properties or render a material functional \cite{Takenaka200587,Takenaka200892,Lin2018152,Iikubo200877,Kodama201081,Lin2015106,Zhang2014115,Lin2012101,Li2011323,Lin2014116,Shao2015396,Born2018749,Wang2011323,Sun201093,Wang200995,Sun200942}.

Among these geometrically frustrated compounds that adopt the cubic (\emph{Space group: Pm$\bar3$m}) structure, antiperovskites Mn$_{3} $GaC and Mn$_{3}$SnC exhibiting first order, volume discontinuous magnetostructural transitions (at $T_{ms}$) have been studied for their giant magnetocaloric properties \cite{Cakir2012100,Wang200985}. While both compounds are known to exhibit ferromagnetic (FM) as well as antiferromagnetic (AFM) orders, a remarkable contrast observed in their magnetostructural properties \cite{Fruchart197844} extended the interest in these materials. Recent reports suggest that the discrepancies basically arise from distortions that are restricted to the local structure surrounding the Mn$_6$C octahedra \cite{Dias2017122,Dias201548}. In particular, x-ray absorption fine structure spectroscopy (XAFS) study on Mn$_3$GaC illustrated a splitting of the Mn--Mn correlation into long and short bond distances. While the longer Mn--Mn bond distances aid the ferromagnetic transition ($T_C$ = 242 K) from a room temperature paramagnetic (PM) state in Mn$_3$GaC, an abrupt decrease in the shorter Mn--Mn distances at $T_{ms}$ = 178 K explains the first order magnetostructural transition to the AFM ground state described by propagation vector $k$ = [$ \frac{1}{2},\frac{1}{2},\frac{1}{2}$]\cite{Fruchart19708,Dias2017122}.

On the other hand Mn$_3$SnC, though it exhibits a single volume discontinuous transition from the PM state to a state with complex magnetic order at about $T_{ms} \sim$ 279 K, time dependent magnetization measurements at the first order transition temperature illustrate an early development of FM order on all three Mn spins along the 001 direction followed by a flipping of spins of two of the Mn atoms to give an additional AFM order at a later time \cite{Cakir201796}. Local structure reports have demonstrated that the Mn$ _{6}$C octahedron in Mn$_3$SnC elongates along one direction and shrinks along the other two while preserving the overall cubic symmetry of the unit cell. As a result, the magnetic propagation vector changes to $k$ = [$ \frac{1}{2},\frac{1}{2},0 $] and allows the existence of AFM order along with a weak FM component \cite{Heritier197712,Dias201548}.

Such a difference in properties of isostructural Mn$_3$GaC and Mn$_3$SnC can be attributed to the size of A-site atom (Ga and Sn). Larger Sn atoms create a chemical pressure on the Mn$_6$C functional unit to alter the nature of magnetic properties. To further understand the role of  the A-site atom in the magnetostructural transition in such antiperovskites, a study of the evolution of the magnetic transformations was carried out using solid solutions of the type Mn$_3$Ga$_{1-x}$Sn$_x$C, (0 $ \le x \le $ 1) \cite{Dias20141}. Herein, though the unit cell volume of the doped compounds, at room temperature, showed a conformity with Vegard's law, their magnetic and magnetocaloric properties behave differently. Increase in cell volume causes the functional units to experience a negative chemical (tensile) pressure in Mn$_3$GaC whereas a positive chemical (compressive) pressure is experienced by Mn$_6$C octahedra in Mn$_3$SnC with Ga doping. Taking a cue from pressure studies on Mn$_3$GaC and Mn$_3$SnC,  Sn doping in Mn$_3$GaC, should result in a decrease of $T_C$ and an increase in $T_{ms}$ while a decrease in $T_{ms}$ should be observed due to Ga doping in Mn$_3$SnC \cite{Bouchaud196637,Shao2015396,Jing201229}. Though a decrease in $T_C$ is observed upon Sn doping in the Ga-rich compositions ($x \le 0.2$), $T_{ms}$ follows a opposite trend. In the Sn-rich compositions ($x \ge 0.8$) Ga doping results in expected decrease in $T_{ms}$. However, the compounds with intermediate concentrations ($0.41 \le x \le 0.71$) present signatures of more than one type of magnetic order \cite{Dias20141}. Presence of a magnetic phase separation was indeed shown in Mn$_3$Ga$_{0.45}$Sn$_{0.55}$C \cite{Dias201795}. The magnetic phase separation occurs due to different local structures of the Mn atoms residing in Ga-rich and Sn-rich regions. However, there are still some unanswered questions. Why are the magnetic and magnetocaloric properties of phase separated Mn$_3$Ga$_{1-x}$Sn$_x$C compounds completely different from those of Mn$_3$GaC or Mn$_3$SnC? and Is this magnetic phase separation limited only to intermediate doping region or is it over the entire solid solution range? Therefore it is pertinent to study the underlying magnetic structure and local structure around the constituent atoms and correlate them with the magnetic properties of these solid solutions. In the present study, using neutron diffraction and x-ray absorption fine structure techniques, we show that even in low doping region of Sn in Mn$_3$GaC or Ga in Mn$_3$SnC, the local structure around Mn can be differentiated into two types, one similar to that in Mn$_3$GaC and other to Mn$_3$SnC. The packing of different local structures into a single global structure results in tensile/compressive strain in the two local structures. Thus the local phase separation is responsible for the observed magnetic properties of Mn$_3$Ga$_{1-x}$Sn$_x$C over the entire solid solution range.

\section {Experimental}
Synthesis of polycrystalline Mn$_3$Ga$_{1-x}$Sn$_x$C, (0 $ \le x \le $ 1) samples using the solid state reaction technique first involved mixing of stoichiometric weights of Mn, Ga, Sn and C with excess graphite powder (about 15 wt.\%) as described in Ref. [\onlinecite{Dias20141}]. This was followed by pelletising and annealing the resulting mixtures in evacuated quartz tubes at 1073 K and 1150 K \cite{Dias20141}. On cooling to room temperature, the compounds were characterized for their phase formation and purity by x-ray diffraction (XRD) using Cu K$_\alpha$ radiation. Magnetization measurements as a function of temperature were carried out between the 5 K to 300 K temperature range during the zero field cooled (ZFC) and field cooled cooling and warming (FCC and FCW) protocols in an applied field of H = 0.01 T. Neutron diffraction experiments as a function of temperature were performed between 6 K and 300 K on the PD2 neutron powder diffractometer ($\lambda$ = 1.2443 \AA) at the Dhruva Reactor (Bhabha Atomic Research Center, Mumbai, India) on three of the samples, each belonging to (a) Ga-rich region ($x < 0.4$), (b) intermediate region ($0.4 < x  < 0.7$) and (c) Sn-rich region ($x > 0.7$). The diffraction patterns were Rietveld refined using the FULLPROF suite software package \cite{Carvajal1993192,Rodriguez}. In order to extract the magnetic contribution more precisely, neutron diffraction experiments were also performed on E6 diffractometer at BER-II reactor (Helmholtz Zentrum Berlin) using neutrons of 2.4 \AA. Extended x-ray absorption fine structure (EXAFS) spectra were recorded in transmission mode at the Mn (6539 eV), Ga (10367 eV) and Sn (29200 eV) K edges using beamlines BL-9C, NW10A at Photon Factory, Japan and P-65 beamline at PETRA-III synchrotron source, DESY, Hamburg, Germany for a few samples spanning over the entire range of solid solution Mn$_3$Ga$_{1-x}$Sn$_x$C. The data were fitted to theoretically modeled spectra using the Demeter program \cite{Ravel200512}.

\section {Results and Discussion}
\begin{figure}[h]
\begin{center}
\includegraphics[width=\columnwidth]{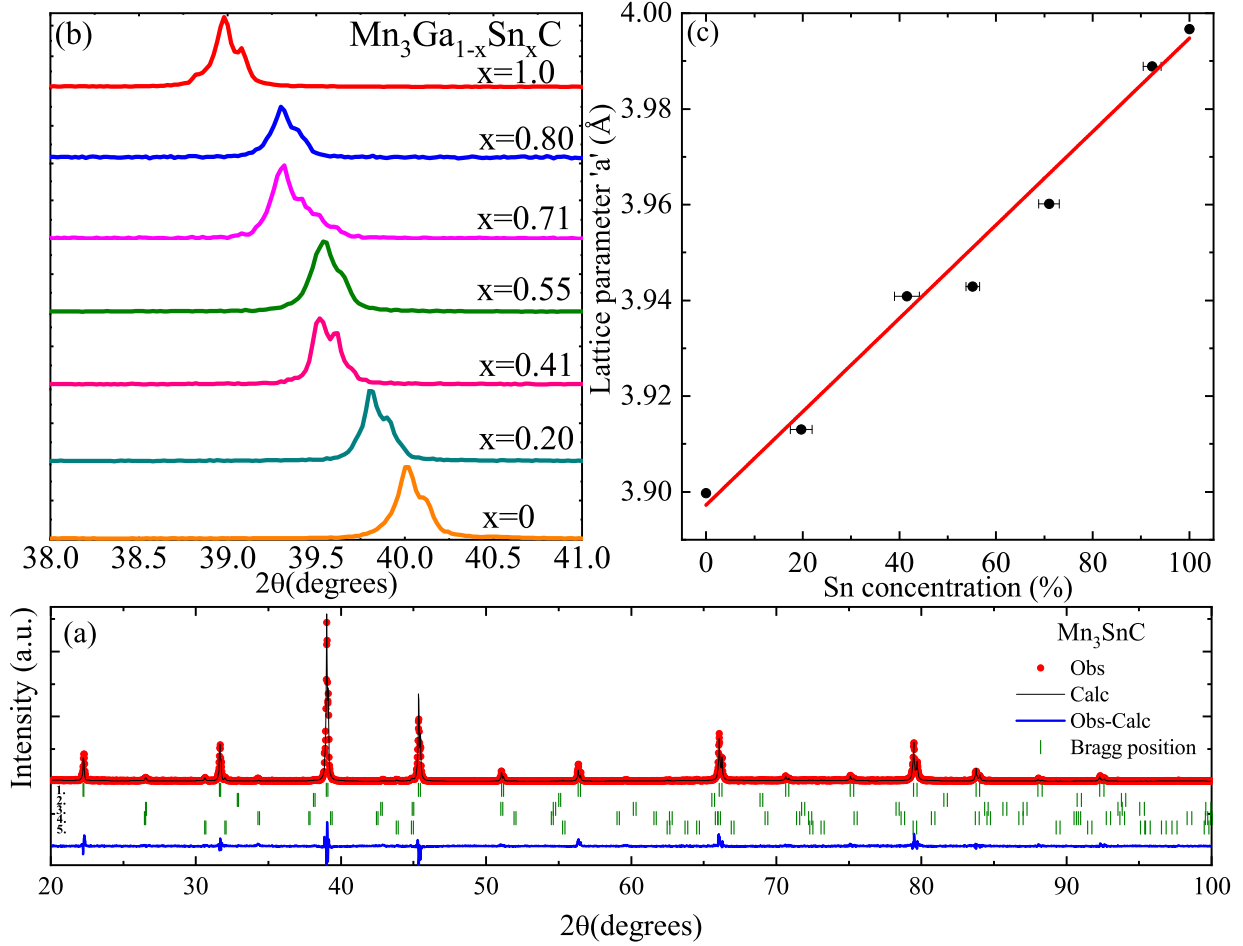}
\caption{(a) Rietveld refined x-ray diffraction pattern recorded for Mn$_3$SnC using Cu K$_\alpha$ radiation at room temperature. Bragg positions from the top correspond to 1. Mn$_3$SnC 2. MnO 3. Graphite 4. SnO and 5. Sn phases respectively. (b) Variation of the (111) x-ray diffraction peak as a function of increasing Sn concentration in Mn$_{3}$Ga$_{1-x}$Sn$_{x}$C, (0 $\leq x \leq$ 1). (c) Systematic increase in lattice constant $a$ depicting the conformity with Vegard's law across the entire composition range.}
\label{xrd}
\end{center}
\end{figure}

\begin{figure}[h]
\begin{center}
\includegraphics[width=\columnwidth]{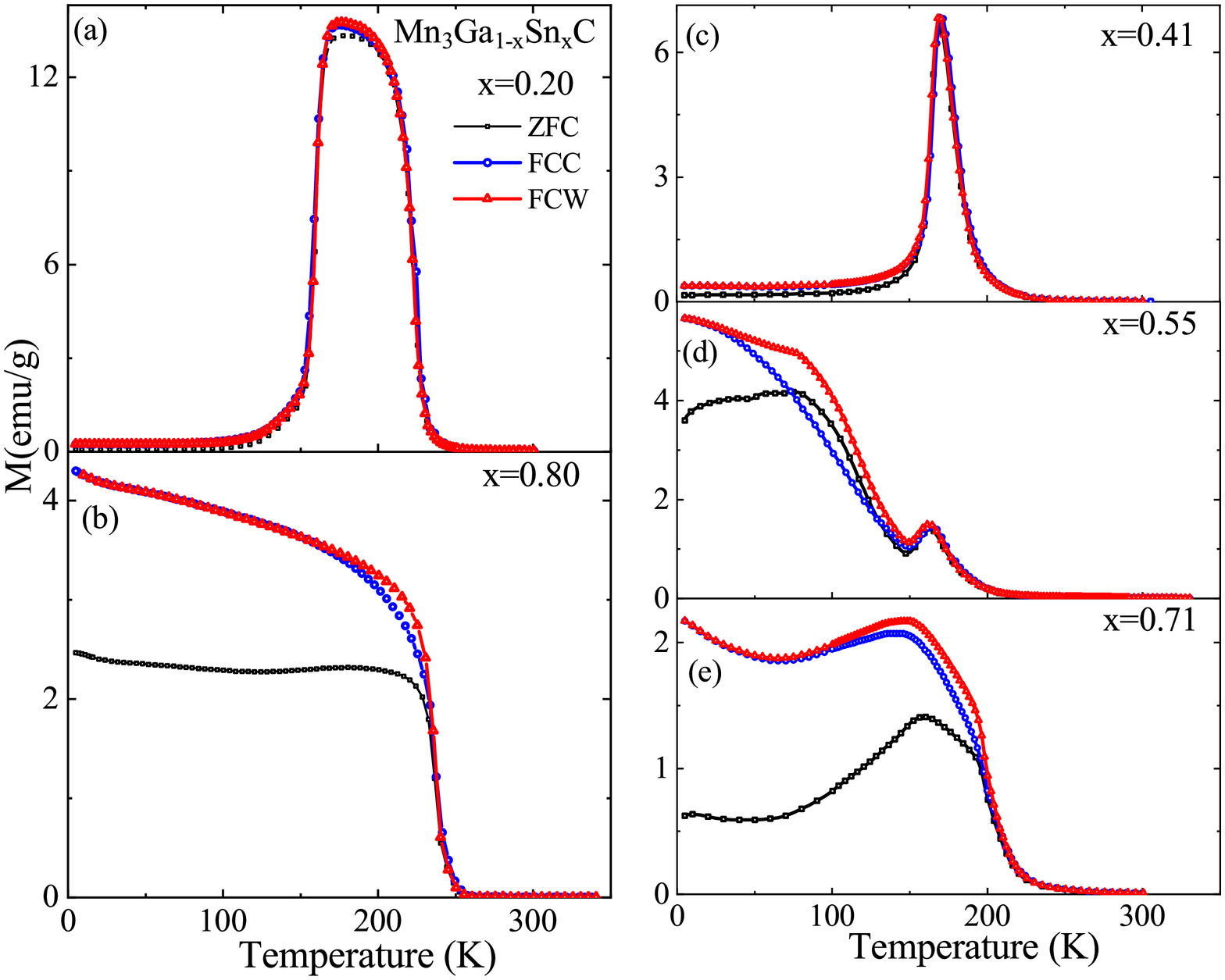}
\caption{Temperature dependence of magnetization measured in 0.01 T applied field during ZFC, FCC and FCW cycles for Mn$_{3}$Ga$_{1-x}$Sn$_{x}$C, (0 $< x <$ 1).}
\label{fig:mtt}
\end{center}
\end{figure}

All the compounds under study crystallized in a cubic phase within the space group $Pm\bar3m$ and have been reported earlier in Ref. \cite{Dias20141}. The antiperovskite structure has Ga or Sn atoms occupying the $1a$ Wyckoff position at the corners of the cube while Mn atoms occupy the face centered $3c$ positions forming a corner sharing Mn$ _{6}$C octahedra with the C atom positioned at the center ($1b$ position) of the cube \cite{Howe19572, Bouchaud196637}. The Rietveld refined diffraction pattern for one of the members of the series, Mn$_3$Ga$_{1-x}$Sn$_x$C ($x$ = 1) is displayed in Fig. \ref{xrd}(a) for clarity. The evolution of maximum intensity peak with changing values of $x$ is shown in Fig. \ref{xrd}(b). A clear shift to higher values of 2$\theta$ can be seen as the value of $x$ decreases from 1 to 0. The corresponding change in the unit cell volume presented in Fig. \ref{xrd}(c) also shows a conformity with Vegard's Law. Such an evolution of lattice volume clearly indicates presence of compressive stress on Mn$_3$SnC or a tensile stress on Mn$_3$GaC. Hence the variation of magnetic properties with doping should have been similar to those observed under pressure. However, the temperature dependent magnetization measurements in Fig. \ref{fig:mtt} present a completely different picture.

It can be seen that the Ga-rich compound, ($x$ = 0.2), has magnetization similar to Mn$_3$GaC (Fig. \ref{fig:mtt}(a)) while the Sn-rich compound, ($x$ = 0.8) exhibits similarity with magnetization behavior of Mn$_3$SnC (Fig. \ref{fig:mtt}(b)). The intermediate compositions, (0.41 $\leq x \leq$ 0.71) whose magnetization behaviors are presented in Figs. \ref{fig:mtt}(c), \ref{fig:mtt}(d) and \ref{fig:mtt}(e) highlight the coexistence of two magnetic phases.

\begin{figure}[h]
\begin{center}
\includegraphics[width=\columnwidth]{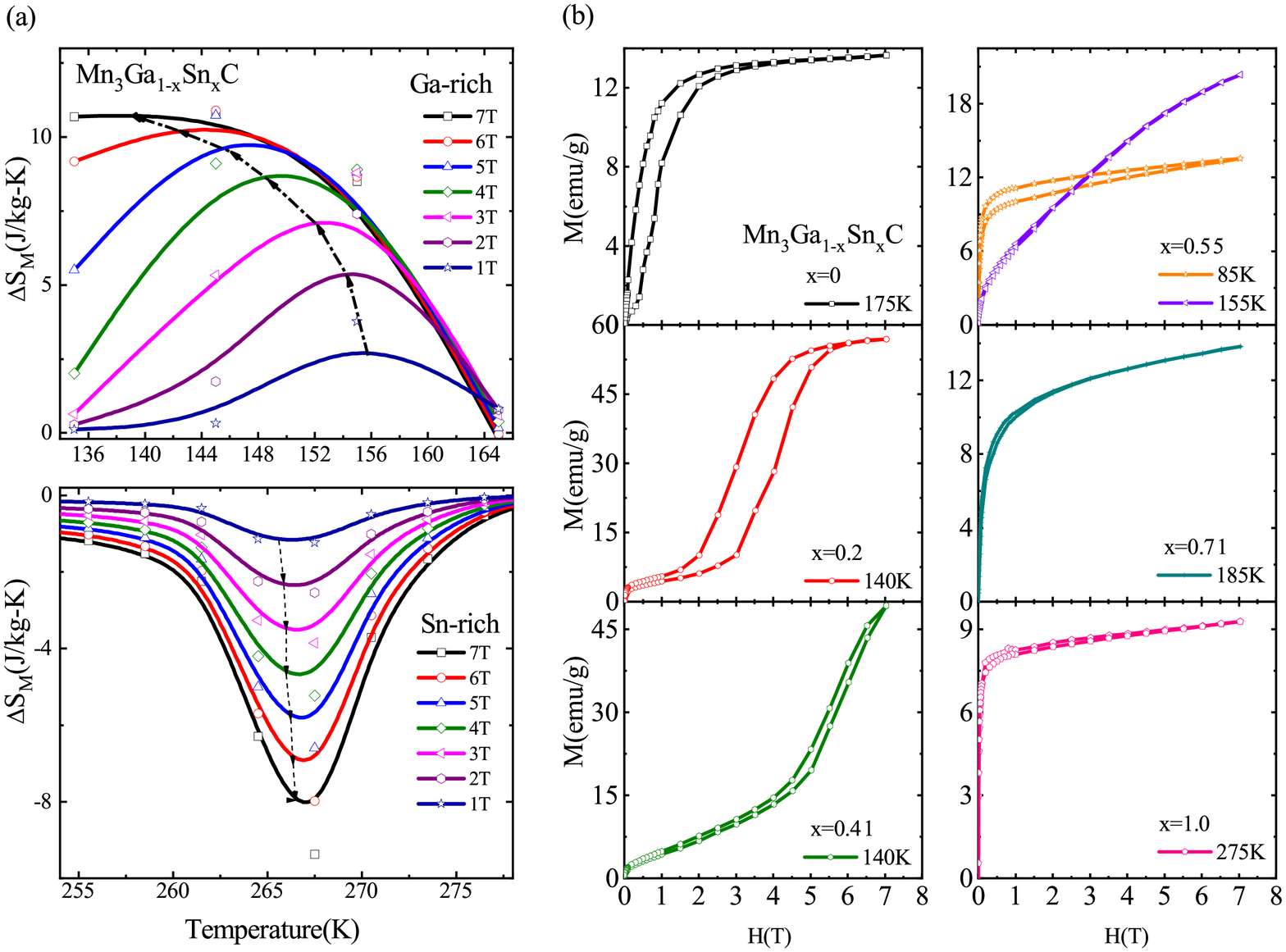}
\caption{(a) Prototype changes in entropy for the Ga-rich and Sn-rich compositions around the magnetostructural transition temperature. (b) Hysteresis loops summarizing the effective strains on the Mn$_6$C octahedra of all Mn$_3$Ga$_{1-x}$Sn$_x$C (0 $\leq x \leq$ 1) compositions at the same (T/T$_N$ $\sim$0.95) relative temperature.}
\label{fig:mcemh}
\end{center}
\end{figure}

\begin{figure}[h]
\begin{center}
\includegraphics[width=\columnwidth]{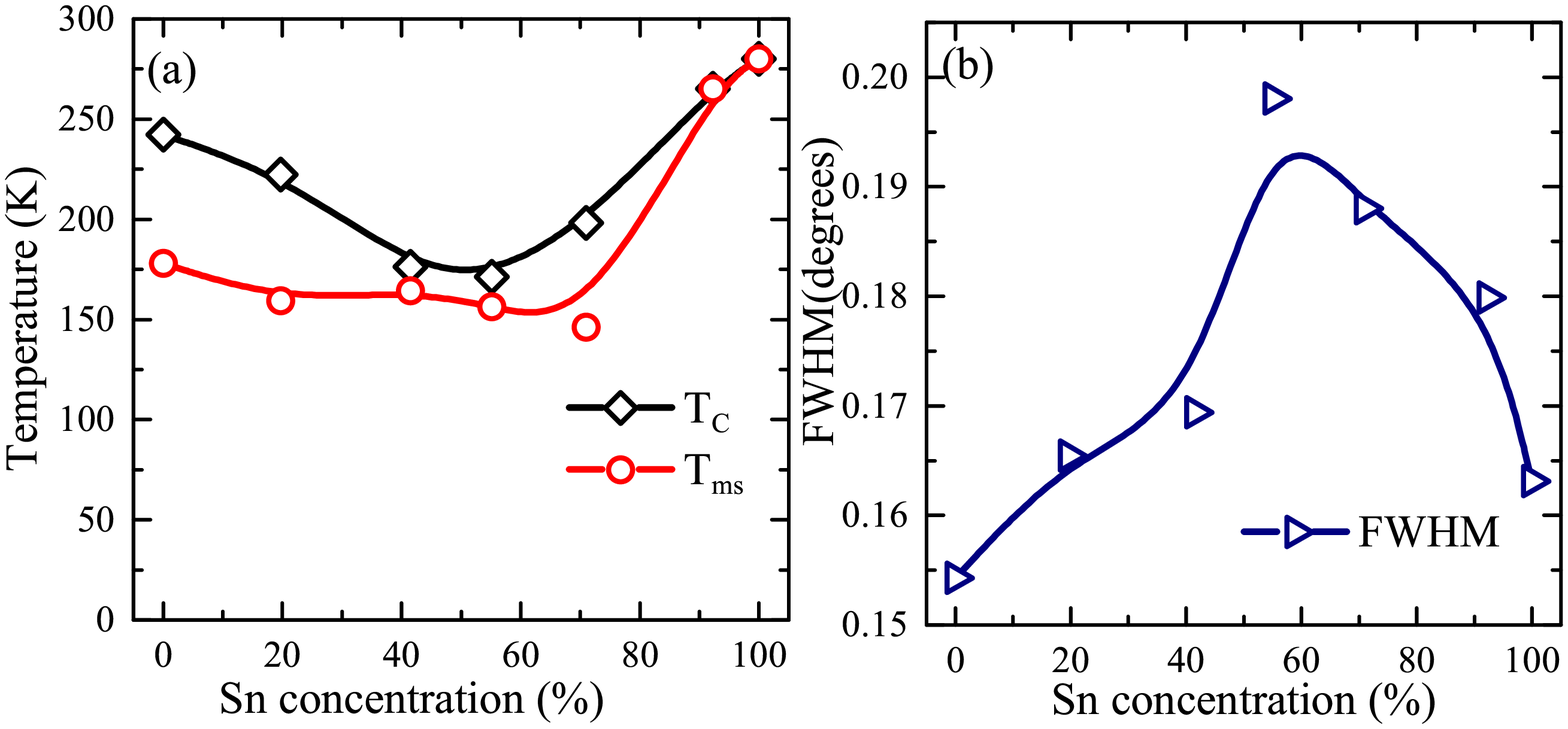}
\caption{Variation of (a) magnetic transition temperature and (b) lattice strain as a function of dopant concentration in Mn$_3$Ga$_{1-x}$Sn$_x$C at 300 K.}
\label{fig:mphase}
\end{center}
\end{figure}

Further, as observed in Fig. \ref{fig:mcemh}(a) the magnetocaloric properties of Mn$_3$Ga$_{1-x}$Sn$_x$C also show a change in character with addition of Sn for Ga. Detailed analysis in Ref. [\onlinecite{Dias2015117}] attributes these changes to strain on the Mn$_6$C octahedra due to comparatively larger Sn in place of Ga atom at the A-site. This strain is also believed to affect the metamagnetic transition seen in Mn$_3$GaC \cite{Dias2014363,Dias2017122}. With increasing Sn doping, Fig. \ref{fig:mcemh}(b)illustrates requirement of a higher and higher magnetic field to induce the metamagnetic transition indicating increase in stiffness of the Mn$_6$C octahedra with increasing Sn doping \cite{Dias2015117}. The stiffness will affect the ferromagnetic and antiferromagnetic interactions present in these Sn doped Mn$_3$GaC compounds. The values of the magnetic transition temperatures are plotted in Fig. \ref{fig:mphase}(a). It can be seen that with Sn doping in Mn$_3$GaC, $T_C$ decreases rapidly, while the first order antiferromagnetic transition temperature, $T_{ms}$ shows only a marginal decrease from 175 K in $x$ = 0 to 155 K in $x$ = 0.55.  On the Sn-rich side, $T_{ms}$ decreases quite rapidly (about 30 K per 10\% change in concentration) with increasing Ga substitution. It has been shown earlier that Mn$_3$Ga$_{0.45}$Sn$_{0.55}$C presents a magnetically phase separated ground state consisting of Ga-rich and Sn-rich regions \cite{Dias201795}. Despite such a ground state, its magnetostructural transition temperature of 155 K is quite different from that of Mn$_3$GaC ($T_{ms}$ = 175 K) and Mn$_3$SnC ($T_{ms}$ = 279 K) due to strain experienced by these Ga-rich and Sn-rich regions. Presence of strain is because Mn$_3$Ga$_{0.45}$Sn$_{0.55}$C has a single phase cubic crystal structure with lattice parameter distinctly different from  Mn$_3$GaC and Mn$_3$SnC.  Such a lattice strain is present across the entire solid solution range is confirmed from the plot of width (FWHM) of main diffraction peak as a function of Sn content in Fig. \ref{fig:mphase}(b). It can be seen that the width of the diffraction peaks of the two end members is minimum and increase with doping on the either side and reaches a maximum value around $x$ = 0.6.  To understand the cause of this lattice strain as well as its effect on long range magnetic ordering, XAFS and neutron diffraction studies on some of the members of this series have been carried out.

Neutron diffraction patterns at several temperatures in the range 6 K to 300 K were recorded and analyzed for three representative compounds, each belonging to the three regions, Ga-rich ($x$ = 0.2), Sn-rich ($x$ = 0.8) and intermediate region ($x$ = 0.55) using neutrons of wavelength $\lambda$ = 1.2443 \AA. Additionally in case of $x$ = 0.55 compound, neutron diffraction patterns were also recorded using longer wavelength neutrons ($\lambda$ = 2.4 \AA) in the temperature range of 2 K -- 300 K and especially with a close temperature interval between 2 K -- 50 K. Rietveld refinement of 300 K neutron diffraction patterns belonging to all the three compounds confirmed the structure to be cubic with space group $Pm\bar3m$ and yielded lattice parameters similar to those obtained from x-ray diffraction experiments.

\begin{figure}[h]
\begin{center}
\includegraphics[width=\columnwidth]{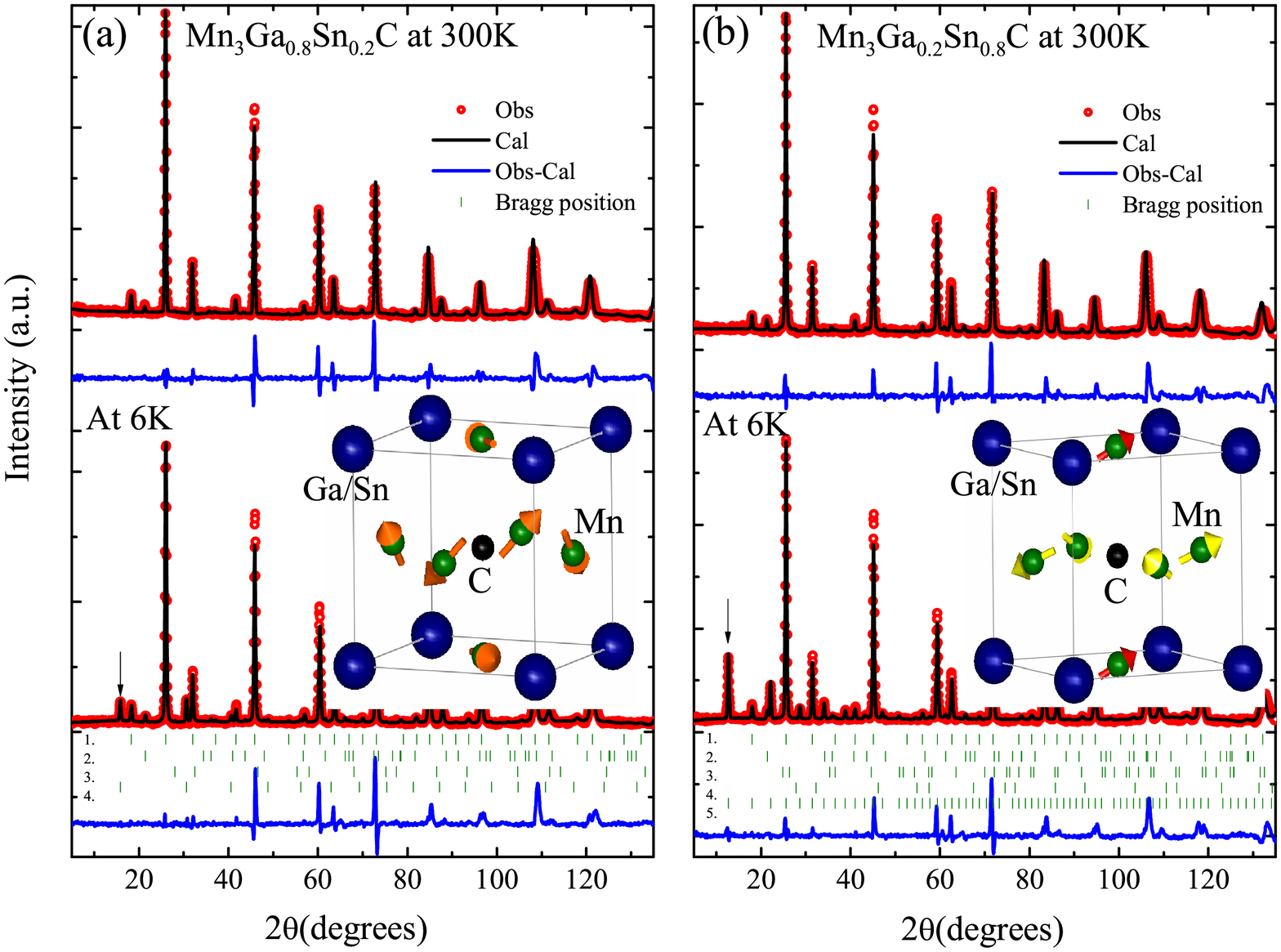}
\caption{Neutron diffraction patterns refined for the (a) Ga-rich and (b) Sn-rich compounds at 300 K and 6 K using the model discussed in the text. Bragg positions for the chemical unit cells are labeled as 1. and associated magnetic reflections are labeled as 4. for Mn$_{3}$Ga$_{0.8}$Sn$_{0.2}$C and 5. for Mn$_{3}$Ga$_{0.2}$Sn$_{0.8}$C. Other positions correspond to minor impurities of Graphite, MnO, and Sn. The insets show the corresponding Mn spin alignment drawn within the chemical unit cell.}
\label{fig:nd1}
\end{center}
\end{figure}

\begin{figure}[h]
\begin{center}
\includegraphics[width=\columnwidth]{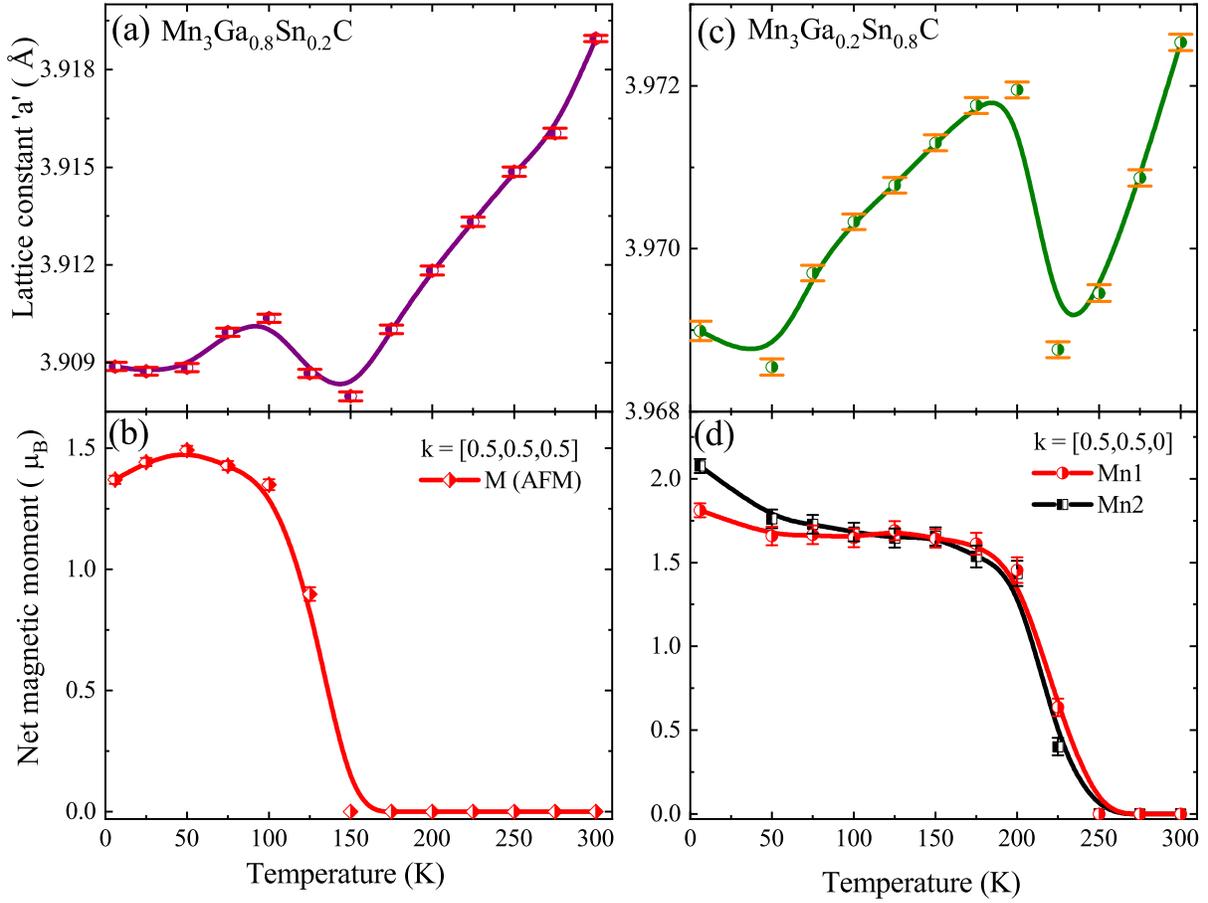}
\caption{A comparison between (a) and (c) thermal variation of refined values of lattice parameter $a$ and (b) and (d) the corresponding values of magnetic moment obtained from the Rietveld analysis of temperature dependent neutron diffraction patterns recorded for compounds Mn$_{3}$Ga$_{0.8}$Sn$_{0.2}$C and Mn$_{3}$Ga$_{0.2}$Sn$_{0.8}$C.}
\label{fig:nd2}
\end{center}
\end{figure}

A comparison between 300 K and 6 K diffraction patterns of $x$ = 0.2 and $x$ = 0.8 compounds in Fig. \ref{fig:nd1} emphasizes on the presence of magnetic reflections that are associated with magnetic structures developing below their respective ordering temperatures. Fixing the magnetic reflections to an absolute scale determined by nuclear reflections in the chemical unit cell at 300 K, a systematic study of the thermal evolution of lattice parameters and magnetic moments was carried out for both the compounds. Predictably, the search for a suitable magnetic propagation vector ($k$ vector) and corresponding magnetic model that results in an excellent fit to the strongest magnetic Bragg peak (2$\theta$ = 15.8$^\circ$) in the Ga-rich sample at 6 K gave rise to the magnetic structure pictorially represented as an inset in Fig. \ref{fig:nd1}(a). Clearly, the structure generated by propagation vector $k = [\frac{1}{2},\frac{1}{2},\frac{1}{2}]$ having ferromagnetic layers stacked antiferromagnetically along the [111] direction and a net AFM moment of 1.37 $ \pm$ 0.02 $ \mu_{B}$ per Mn atom is comparable with the structure reported for Mn$  _{3}$GaC in zero applied field \cite{Fruchart197844,Cakir2014115}. Effect of doped Sn atom is mainly seen on the orientation of Mn spins. Compared to Mn$_3$GaC, wherein Mn moments have all three, M$_x$, M$_y$ and M$_z$ components, in $x$ = 0.2 compound, the Mn moments are oriented within the respective planes and thus have only two of the three magnetic components. The thermal variations in refined lattice parameter $a$ and ordered magnetic moment presented in Fig. \ref{fig:nd2}(a) highlight the existence of a magnetostructural transition. With decreasing temperature, the lattice constant decreases linearly down to 150 K with a slight change of slope near $T$ = 250 K marking the onset of a transition from the PM--FM state. With further decrease in temperature, an abrupt increase in lattice parameter ($\sim$ 0.06\%) accompanied by simultaneous appearance of the ordered AFM moment in Fig. \ref{fig:nd2}(b) illustrates the reported \cite{Dias20141} first order transition from a low volume FM state to a higher volume AFM state. Once the AFM phase is established, the lattice parameter shows no further variation with decrease in temperature.

A similar approach of adopting a larger magnetic unit cell generated by the propagation vector $k = [\frac{1}{2},\frac{1}{2},0]$ that describes the magnetic structure of Mn$ _{3} $SnC resulted in a good fit to the most intense magnetic Bragg peak (2$\theta$ =12.7$^\circ$) of the Sn-rich sample. The complex magnetic structure obtained at 6 K and depicted within the chemical unit cell in the inset of Fig. \ref{fig:nd1}(b) has spins on two of the Mn (indicated by yellow arrows) atoms arranged in a square plane to give a net antiferromagnetic moment of 2.08 $ \pm $ 0.04 $ \mu_{B}$. The spin on the third Mn atom (indicated by red arrows) has a slightly canted FM alignment in the crystallographic unit cell and a net moment of 1.81 $ \pm$ 0.04 $ \mu_{B}$. Here there are significant changes not only in the orientation of Mn moments but also in their magnitudes. In Mn$_3$SnC, the Mn(2) moments were ferromagnetic and oriented along the 001 direction with a magnitude of 0.7 $\mu_B$, while in the case of Mn$_3$Sn$_{0.8}$Ga$_{0.2}$C they are tilted away from the $z$ axis resulting in canted arrangement along the $[\frac{1}{2},\frac{1}{2},0]$ direction, resulting in a net ferromagnetic moment of $0.26 \pm 0.04 \mu_B$ at 6 K. This net ferromagnetic moment also explains the rise in magnetization below $T_{ms}$ (see Fig. \ref{fig:mtt}(a)) whereas the compressive strain produced by smaller Ga leads to a change in spin alignment. The thermal evolution of the Sn-rich cell parameter is very similar to that of Mn$_{3} $SnC. Fig. \ref{fig:nd2}(c) shows that the monotonous decrease in $a$ down to 225 K is followed by an abrupt increase of about 0.08\%, characteristic of the first order magnetic transition from the PM to a complex magnetically ordered ground state. Magnetic moments for both Mn1 and Mn2 atoms also appear at the same temperature as the lattice anomaly as shown in Fig. \ref{fig:nd2}(d).

\begin{figure}[h]
\begin{center}
\includegraphics[width=\columnwidth]{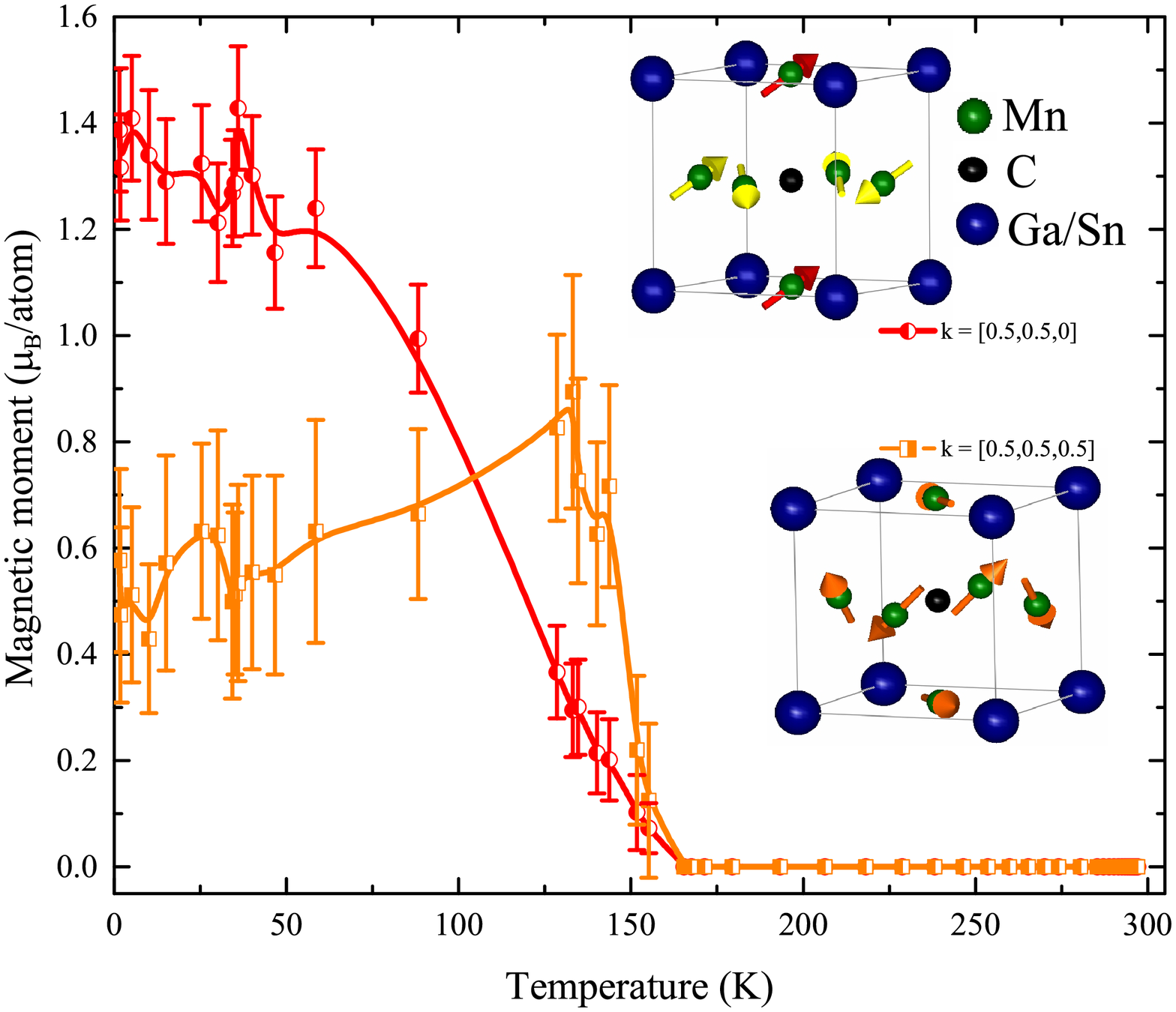}
\caption{Magnetic moment variation calculated from neutron diffraction patterns ($\lambda$ = 2.4 \AA) in the 2 K -- 175 K range for $k = [\frac{1}{2},\frac{1}{2},\frac{1}{2}]$ and $k = [\frac{1}{2},\frac{1}{2},0]$.}
\label{fig:nd4}
\end{center}
\end{figure}

Neutron diffraction measurements on $x$ = 0.55 though were reported earlier \cite{Dias201795}, have been repeated again on a powder diffractometer with a better resolution (E6 at HMI, Berlin) and at a much closer temperature interval especially in the temperature range from 2 K to 50 K as 30 K was identified as the cluster glass transition  temperature \cite{Dias201795}. The need of second set of measurements on $x$ = 0.55 arose in order to understand its completely different bulk magnetic properties compared to the two end members, despite sustaining two independent magnetic spin alignments identical to that in Mn$_3$GaC and Mn$_3$SnC respectively. However, not much difference was noticed between the two neutron diffraction studies performed at Dhruva earlier \cite{Dias201795} and HMI Berlin. Two long range AFM orders, described by propagation vectors $k = [\frac{1}{2},\frac{1}{2},\frac{1}{2}]$   and $k = [\frac{1}{2},\frac{1}{2},0]$ respectively appear at the same temperature ($T_{ms}$ = 155 K) concomitant with abrupt change in lattice volume.   The magnetic moments estimated from these measurements are presented in Fig. \ref{fig:nd4} which is more or less similar to that reported earlier \cite{Dias201795}.  Magnetic moment evolution of the two magnetic phases also do not show any significant change from the previous measurements and especially around the reported cluster glass transition temperature of $T$ = 30 K. This indicates that the dip in magnetization obtained in bulk magnetic measurements is indeed due to a glassy transition and not due to any magnetic interaction between the two magnetic phases.

\begin{figure}[h]
\begin{center}
\includegraphics[width=\columnwidth]{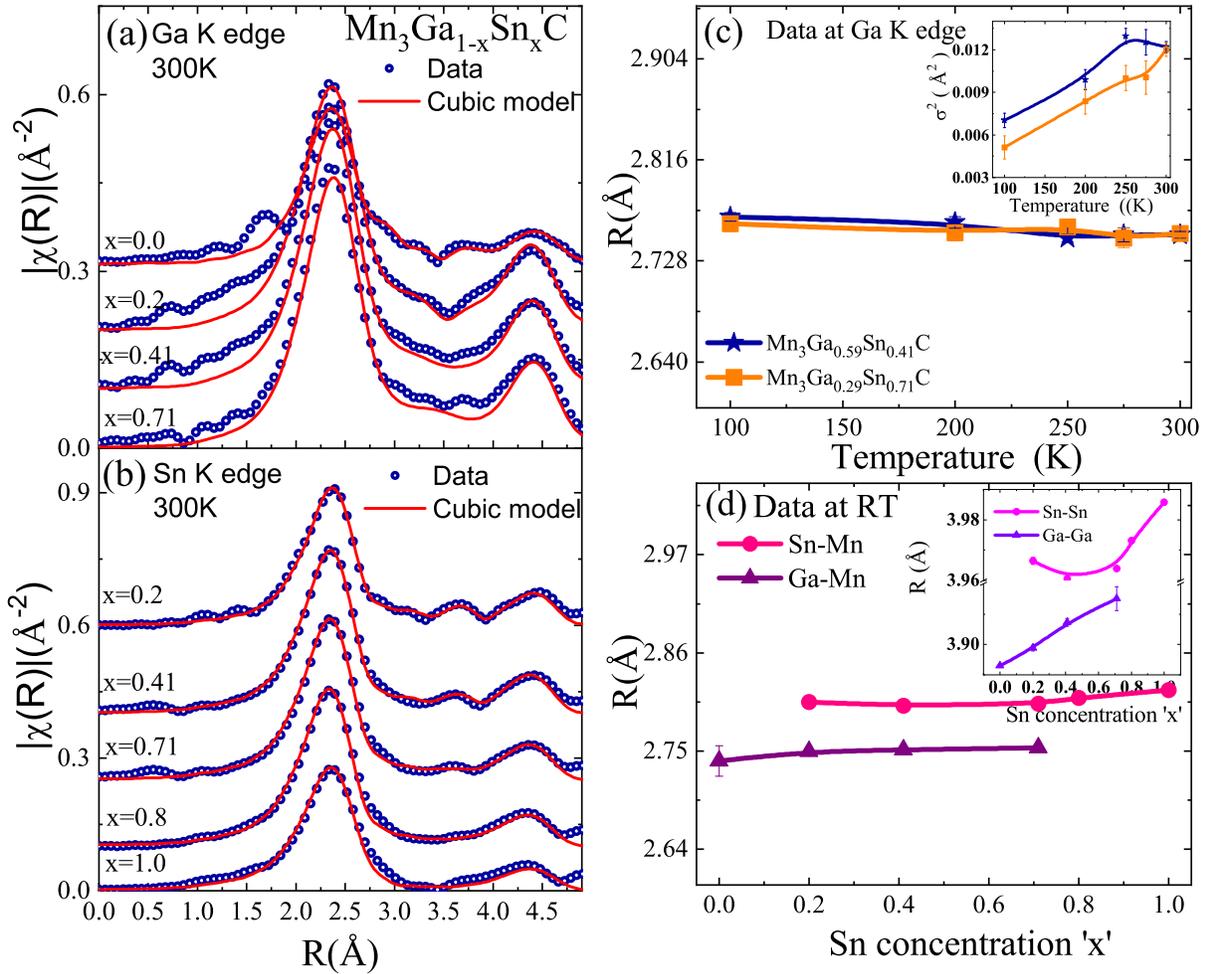}
\caption{(a) and (b) FT of the Ga and Sn K edge XAFS data for 0.2 $\leq x \leq$ 1 compounds. The plots show the perfect fit to the cubic model at 300 K. (c) Thermal variation of the Ga--Mn bond lengths obtained from a series of temperature dependent XAFS data recorded at the Ga K edge. (d) Room temperature variation of the Ga--Mn and Sn--Mn bond lengths as a function of increasing Sn concentration. Inset shows the room temperature variation of the Ga--Ga and Sn--Sn bond distances as a function of Sn concentration.}
\label{fig:gasn}
\end{center}
\end{figure}

Despite solving magnetic structures, the exact cause of strain and its effect on the magnetic properties of the doped compounds, especially those belonging to the intermediate region is still unclear. Since all compounds crystallize in a single cubic phase implying long range order, the key to understanding strain and its effect on magnetic properties could be in studying the local structures around the metal atoms. Therefore to investigate the role of short range interactions in these compounds, temperature dependent EXAFS spectra recorded at the Ga, Sn and Mn K edges in few select compounds were independently analyzed. At first, contributions from the various near neighbor atoms at different distances were isolated by Fourier transforming the weighted XAFS signal from $k$ to $R$ space. While considering Ga or Sn as the absorbing atom, scattering from the eight nearest Mn atoms in the first coordination shell at bond distance R$'$ $\sim$ 2.8 \AA~ is well separated from all other correlations (Ga--C/Sn--C at R$'$ $\sim$ 3.38 \AA~ and Ga--Ga/Sn--Sn at R$'$ $\sim$ 3.90 \AA) and solely contribute to the main peak centered around phase uncorrected radial distance R $\sim$ 2.5 \AA~ in Figs. \ref{fig:gasn} (a) and \ref{fig:gasn} (b). Analysis of Ga or Sn K XAFS are effective in unambiguously providing Ga/Sn--Mn bond distance.

The analysis of Ga and Sn K XAFS indicates two interesting aspects. Firstly, the cubic long range order requires Ga--Mn and Sn--Mn bond distances to be equal in a given compound. Secondly, the Vegard's law behavior of lattice constants demands an increase in Ga--Mn or Sn--Mn distance with $x$. However, the obtained Ga--Mn distances for $x$ = 0.2, 0.41 and 0.71 are nearly equal to Ga--Mn distance in Mn$_3$GaC (2.74 \AA) as illustrated in Fig. \ref{fig:gasn}(d). Similarly the Sn--Mn distance in all the doped compounds studied here is $\sim$ 2.81 \AA~ which is equal to the Sn--Mn distance in Mn$_3$SnC. From the context of XAFS, such a situation suggests that local symmetry is maintained even though the compounds possess long range structural order. Another interesting aspect is that the Ga-- Mn and Sn--Mn bond distances show negligible temperature dependence across the magnetostructural transition but the mean-square relative displacement ($\sigma^{2}$) in these distances do show an abrupt increase at the first order transitions as can be seen in the inset of Fig. \ref{fig:gasn}(c).

Further the nonequivalence of Ga--Mn and Sn--Mn bond distances in the doped compounds is also seen in Ga--Ga and Sn--Sn distances. The inset of Fig \ref{fig:gasn}(d) depicts variation of Ga--Ga and Sn--Sn distances in all doped compounds along with Mn$_3$GaC and Mn$_3$SnC. The variation of Ga--Ga and Sn--Sn distances reflect two important aspects. First, Ga--Ga or Sn--Sn distance is the same as lattice constant and hence should have been identical to each other and to the respective lattice constants of doped compounds. Their non equivalence hints towards a possibility of existence of Ga-rich and Sn-rich regions in all doped compounds. Second, their variation with $x$ suggests that in all the doped compounds, these Ga-rich regions and Sn-rich regions try to match their lattice constants with each other in order to preserve the long range structural order. The apparent failure of the two regions to match their lattice constants as indicated by non equivalent Ga--Ga and Sn--Sn distances causes strain and it increases as one moves away from the two undoped end members.

Furthermore information is revealed from the analysis of Mn K edge XAFS. It comprises of contribution from two nearest neighbor carbon atoms to the weak first peak in the Fourier transform of XAFS spectra. The second and the strongest peak has contributions of scattering from eight Mn and four Ga or Sn atoms which have the identical bond length. In any of the doped  Mn$_{3}$Ga$_{1-x}$Sn$_{x}$C (0 $< x < $ 1) compounds, the second peak therefore, will have contributions from Mn--Ga, Mn--Sn as well as Mn--Mn scattering. This makes number of fitting parameters quite large. Therefore, an attempt to understand the type of distortions present in these compounds was first made by utilizing the linear combination fitting (LCF) subroutine available in the ATHENA software package \cite{Ravel200512}. The method involves the modeling of a spectrum (LCF fit) as a sum of the constituent spectra of more than one model compounds. In the present case, experimental spectra of all doped compounds were attempted to be described as the sum of experimental spectra of Mn$_3$GaC and Mn$_3$SnC.

\begin{figure}[h]
\begin{center}
\includegraphics[width=\columnwidth]{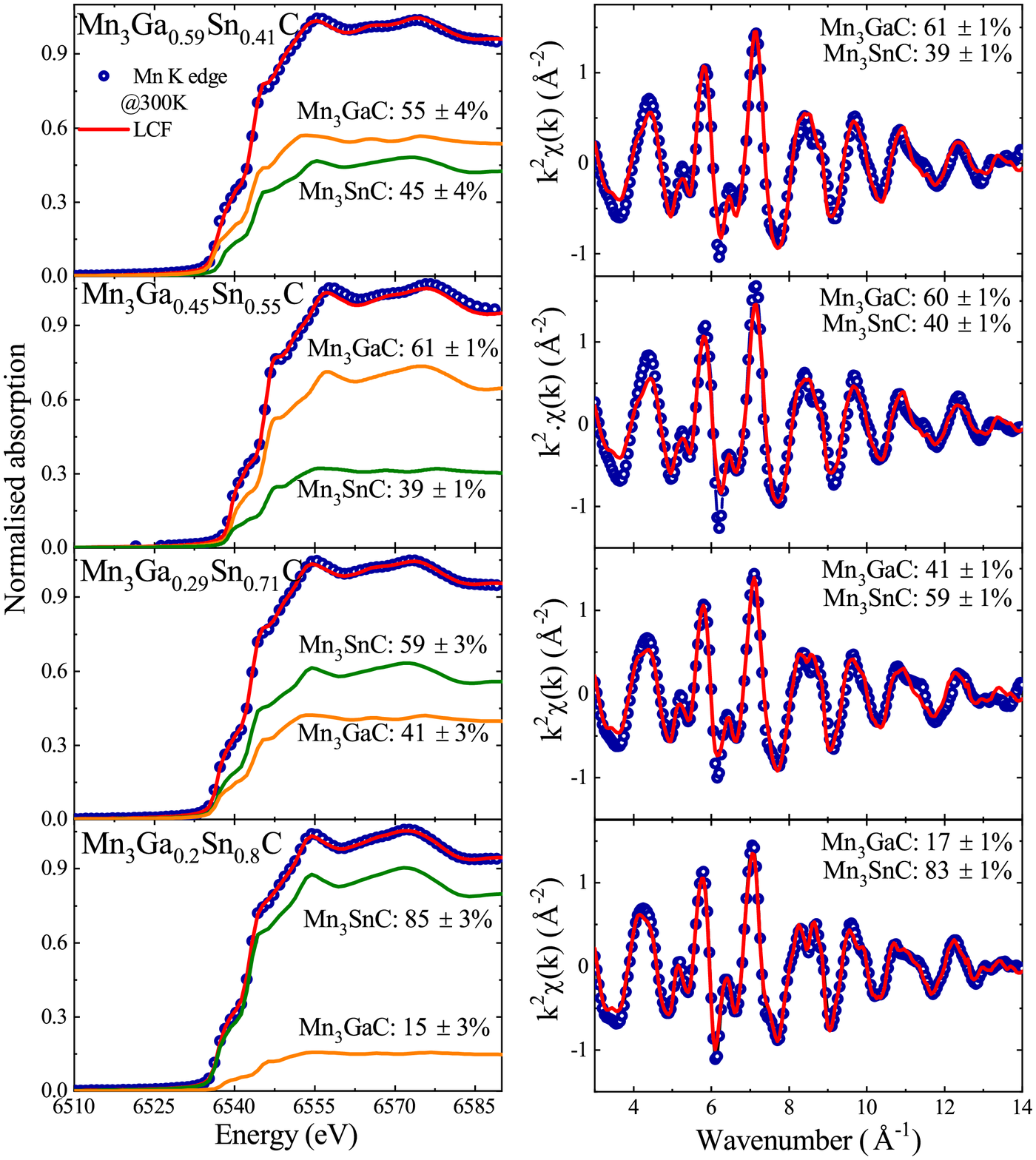}
\caption{Linear combination fitting (LCF) analysis of the XANES and XAFS spectra recorded for all compounds in the -20 to +30 eV and the 3.0 -- 14.0 \AA$^{-1}$ $k$ range respectively. Orange and green solid lines represent contributions from the Mn$ _{3} $GaC and Mn$ _{3} $SnC spectra respectively.}
\label{fig:lcf}
\end{center}
\end{figure}

The resulting fittings to room temperature XANES spectra described in the -20 to +30 eV (with respect to the edge energy) energy range as well as XAFS spectra in the $k$ range of 3.0 -- 14.0 \AA$^{-1}$ are shown in Fig. \ref{fig:lcf}. A perfect fit to both the XANES and $k^2$ weighted XAFS signals recorded for all compositions using a linear combination ratio that matches the starting compositions gives further evidence to the presence of Ga-rich and Sn-rich clusters in varying proportion over the entire solid solution range.

\begin{figure}[h]
\begin{center}
\includegraphics[width=\columnwidth]{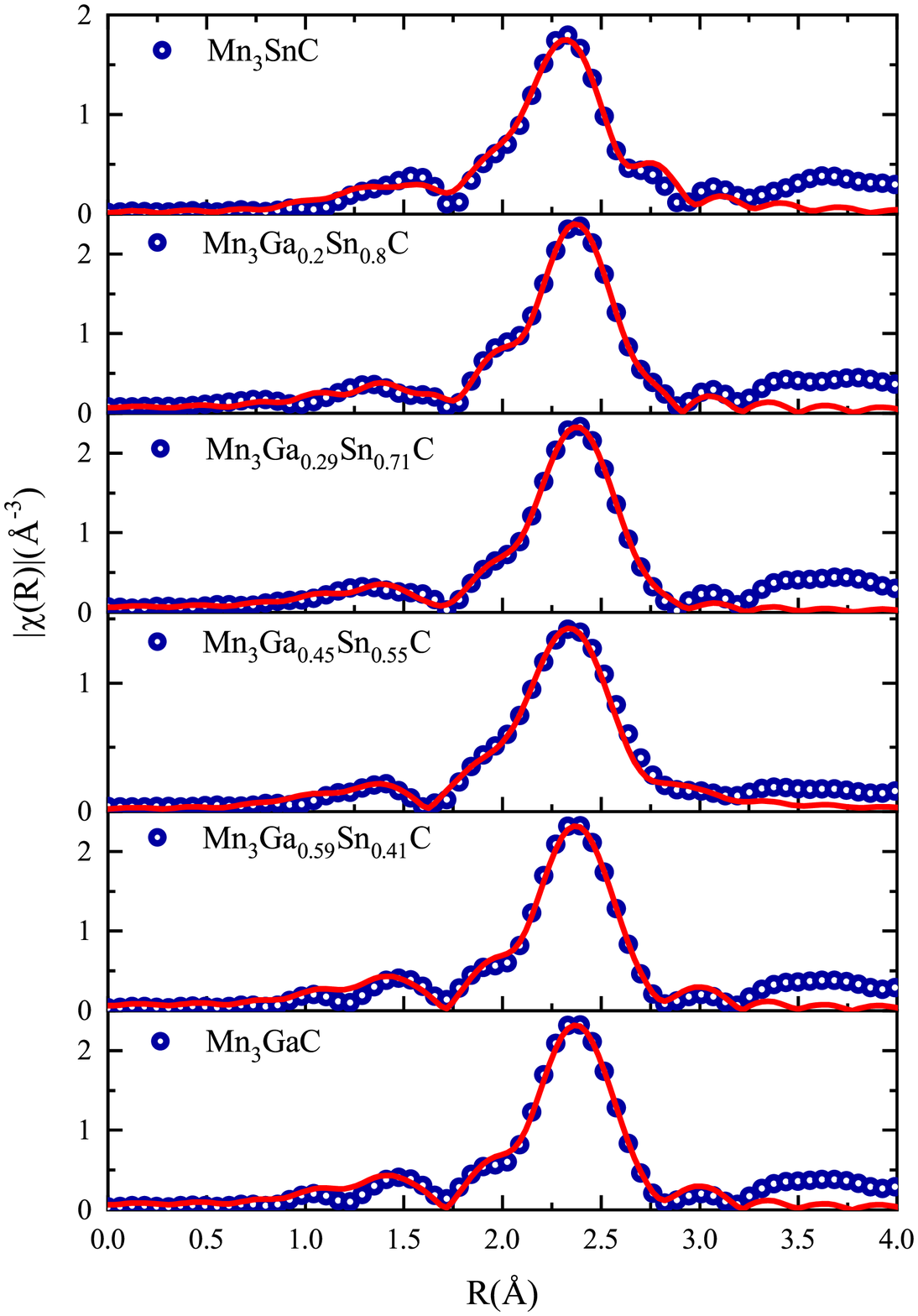}
\caption{100 K XAFS data recorded at the Mn K edge of Mn$_{3}$Ga$_{(1-x)}$Sn$_{x}$C (0 $\leq x \leq$ 1) compounds.}
\label{fig:xafs1}
\end{center}
\end{figure}

With the help of Ga--Mn and Sn--Mn distances obtained from Ga and Sn XAFS analysis respectively, all important Mn--Mn distances were obtained from fitting the Mn EXAFS up to 3 \AA~ in $R$ space. Fittings to data taken at 100 K are shown in Fig. \ref{fig:xafs1}. A good fit was obtained at all temperatures for all the compounds. XAFS analysis of Mn K edge revealed that the Mn--Mn correlation in all these compounds are split into two, short and long distances.

\begin{figure}[h]
\begin{center}
\includegraphics[width=\columnwidth]{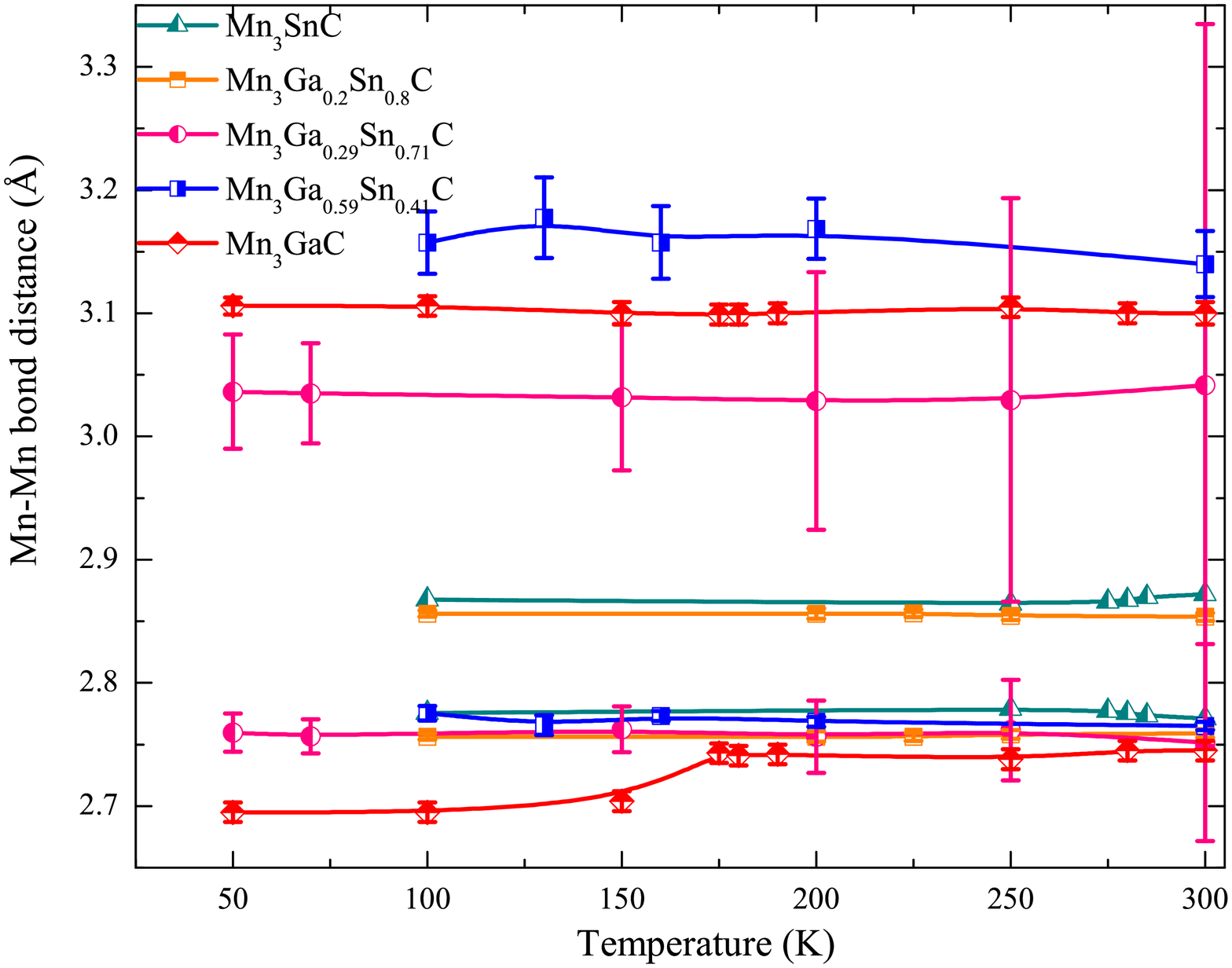}
\caption{Thermal variation of the Mn--Mn bond distances as obtained from the analysis of XAFS data recorded at the Mn K edge of Mn$_{3}$Ga$_{(1-x)}$Sn$_{x}$C (0 $\leq x \leq$ 1) compounds.}
\label{fig:xafs2}
\end{center}
\end{figure}

The separation between Mn--Mn long and short distances across the solid solution as well as their temperature variation is presented in Fig. \ref{fig:xafs2}. The separation between these long and short bond lengths seems to be dependent on effective size of the A-site atom. As can be seen from Fig. \ref{fig:xafs2}, the difference between Mn--Mn$_{short}$ and Mn--Mn$_{long}$ bond distances is largest for Ga-rich compounds (smaller effective size) and it decreases for Sn-rich compounds (bigger effective size). The larger the Mn--Mn$_{long}$ bond distance ($>$ 2.8 \AA), stronger are the ferromagnetic interactions. Similarly smaller the Mn--Mn$_{short}$ distance ($<$ 2.8 \AA) stronger is the antiferromagnetism. Mn$_{3}$GaC which has one of the widest variation of Mn--Mn distances shows both magnetic orders and even in the antiferromagnetic state, ferromagnetism can be induced by application of magnetic field or hydrostatic pressure. In Mn$_{3}$Ga$_{0.59}$Sn$_{0.41}$C compound, Mn--Mn$_{long}$ is longest but at the same time, Mn--Mn$_{short}$ is also quite small and hence it presents a strong competition between ferromagnetic and antiferromagnetic interactions. This compound neither orders ferromagnetically nor antiferromagnetically (Fig. \ref{fig:mtt}(c)) but presents signatures of both orders. Sn-rich compounds ($x >$ 0.7) exhibit comparatively smaller difference between the two Mn--Mn distances and though have a high magnetic ordering temperature, but present a complex magnetic order. Here the two distances are quite close to the critical Mn--Mn distance of 2.8 \AA~ and hence explains the complex magnetic order. The reason for differing distortions in Mn$_{6}$C octahedra can also be understood from this plot. For A-site atom of smaller size like Ga, the Mn$_{6}$C octahedra are less strained and therefore can distort freely resulting in larger difference between the long and short Mn--Mn distances. With increase in the effective size of A-site atom, the octahedra become more and more strained and the distortions are restricted and also result in higher magnetostructural transformation temperature.

\section*{Conclusion}
In conclusion the above studies highlight the importance of the role of A-site atom in magnetostructural transformations in Mn$_{3}$Ga$_{1-x}$Sn$_{x}$C compounds. Replacement of Ga by larger Sn results in formation of Ga-rich or Sn-rich clusters. Although the long range structure is cubic, locally Mn atoms find themselves either in Ga-rich or Sn-rich environments. Mn$_{6}$C octahedra which are surrounded by Sn are tensile strained which affect their magnetic and magnetocaloric properties. Furthermore the formation of Ga-rich and Sn-rich clusters results in a non-ergodic ground state as found in Mn$_3$Ga$_{0.45}$Sn$_{0.55}$C.

\section*{Acknowledgements}
Authors thank Photon Factory, KEK, Japan for beamtime on beamlines 9C and NW10A for the proposal No. 2014G042. Portions of this research were carried out at the light source PETRA III at DESY, a member of the Helmholtz Association (HGF). We would like to thank Edmund Welter and Roman Chernikov for assistance at beamline P65. Financial support by the Department of Science and Technology, (Government of India) provided within the framework of India@DESY collaboration is gratefully acknowledged.

\bibliographystyle{apsrev4-1}
\bibliography{References}

\end{document}